\documentclass{pasj01}
\Received{$\langle$reception date$\rangle$}
\Accepted{$\langle$acception date$\rangle$}
\Published{$\langle$publication date$\rangle$}
\usepackage{amsthm}
\usepackage{deluxetable}
\usepackage{lineno}

\usepackage{color}%, ulem}
\newcommand{\ba}{\begin{eqnarray}}
\newcommand{\ea}{\end{eqnarray}}

\newcommand{\kms}{km s$^{-1}$ }

\begin{document}

\title{High-sensitivity VLBI Observations of the Water Masers in the Seyfert Galaxy NGC 1068 }
\author{Yuna Morishima, \altaffilmark{1}  Hiroshi Sudou,\altaffilmark{1,*},  Aya Yamauchi, \altaffilmark{2} Yoshiaki Taniguchi,\altaffilmark{3}   and Naomasa Nakai\altaffilmark{4} }
\altaffiltext{1}{Faculty of Engineering, Gifu University, 1-1 Yanagido, Gifu, Gifu 501-1193}
\altaffiltext{2}{Mizusawa VLBI Observatory, National Astronomical Observatory of Japan, 2-12 Hoshigaoka-cho, Mizusawa, Oshu, Iwate 023-0861}
\altaffiltext{3}{The Open University of Japan, 2-11 Wakaba, Mihama-ku, Chiba, Chiba 261-8586}
\altaffiltext{4}{Department of Physics and Astronomy, School of Science, Kwansei Gakuin University, 2-1 Gakuen, Sanda, Hyogo 669-1337}
\email{sudou@gifu-u.ac.jp}


\maketitle

\begin{abstract}
We present observational results of water vapor maser emission with our high-sensitivity 22 GHz VLBI imaging of the Seyfert galaxy  NGC 1068.
In this galaxy, there are the following four nuclear radio sources; NE, C, S1, and S2.
Among them, the S1 component has been identified as the nucleus while the C component has been considered as attributed to the radio jet. 
%%%The previous VLBI imaging has detected the water maser emission in both the S1 and the C components. 
In our VLBI observation, we find the following two types of the water maser emission at the S1 component.
One is the  linearly aligned component that is considered as an edge-on disk with the inner radius of 0.62 pc.
The dynamical mass enclosed within the inner radius was estimated to be $1.5\times10^7 M_{\odot}$ by assuming the circular Keplerian motion. 
Note, however, that the best fit rotation curve shows a  sub-Keplerian rotation ($v\propto r^{-0.24\pm0.10}$). 
The other is the water maser emission distributed around the rotating disk component up to 1.5 pc from the S1 component,
%%%The maser emitting regions  in NE tend to be blue-shifted relative to the systemic velocity  ($-$300 \kms) while those in SW tend to be red-shifted  ($+$300 \kms). 
%%%These properties 
suggesting the bipolar outflow from the S1 component.
  { Further, we detected the water maser emission in the C component for the first time with VLBI, and discovered  a ring-like distribution of the water maser emission.}
It is known that a molecular cloud is associated with the C component
  (both  HCN and HCO$^+$ emission lines are detected by  ALMA).
Therefore,  the ring-like maser emission can be explained by the jet collision to the molecular cloud.
However, if these ring-like water masing clouds constitute a rotating ring around the C component, it is likely that the C component also has a supermassive black hole
with  the mass of  $\sim 10^6 M_{\odot}$ that could be supplied from a past minor merger of a nucleated satellite galaxy. 

\end{abstract}

%%%%%\linenumbers

\section{Introduction}

 Some active galactic nuclei (AGNs) emit a large amount of radiation as well as jet phenomena, 
 that are considered to be driven by the gas accretion onto the nucleus of galaxies (e.g., Rees 1984). 
It is now widely accepted that the nucleus of galaxies harbors a  SMBH with a mass of millions of the solar mass or more
 (e.g., Magorrian et al. 1998, Kormendy \& Ho 2013, Genzel 2014). 
The existence of such a SMBH has been proved by observations of the water maser emission exhibiting  a compact disk in a Keplerian rotation around a SMBH
 (e.g., Nakai et al. 1993, Miyoshi et al. 1995, Moran et al. 1995). 
Therefore, thanks to very high-resolution imaging with VLBI,
water maser emission provides a good probe to directly investigate the structure and dynamics of AGNs on a parsec or a sub-parsec scale.

NGC 1068 is one of the archetypical AGN-hosting galaxies.
The systemic velocity of the galaxy is about $V_{\rm LSR}=1137$ \kms (Brinks et al. 1997, Gallimore et al. 2001). 
This galaxy has a compact nuclear radio jet consisting of four components called as S1, S2, C, and NE. 
Among them, the S1 component has been identified as a real nucleus that has a SMBH with mass of $\sim 10^7$ $M_\odot$ (Greenhill et al. 1996; Kormendy \& Ho 2013). 
The evidence of the SMBH has been obtained from the compact water maser emitting region at 22 GHz around the S1 component that was identified 
as a molecular torus surrounding  the SMBH at the center of NGC 1068 (Greenhill et al. 1996, Greenhill \& Gwinn 1997, Gallimore et al. 1997). 
Recent ALMA observations also confirmed the presence of the rotating dusty torus at the nucleus
 (Garcia-Burillo et al. 2016, Gallimore et al. 2016, Imanishi et al. 2016, 2018, Impellizzeri et al. 2019). 
However, another water maser emitting region has been also found in the radio jet component C   {with VLA} (Gallimore et al. 1996a, 2001). 
{Since the possible masing mechanism in this component is kinematical collisional pumping by the jet from the S1 component (e.g., Gallimore 1996a,  Gallimore 1996b), this masing region has been called as the jet maser. }

In this paper, we show our new results of the VLBI imaging of the water maser emission at the components S1 and C in NGC 1068 
by using high-sensitivity array consisting of VLBA, phased-VLA, and   { Effelsberg} 100-m telescope. 

Throughout this paper, we use a distance of NGC 1068, 15.9 Mpc (Kormendy \& Ho 2013).
The adopted basic parameters of NGC 1068 are summarized in table 1. 
All the Velocities used here are in the radio definition and with respect to the Local Standard of Rest (LSR). 

\section{Observations and Data Reduction}

The observations of water masers emission ($J_{\rm K_a K_c}=6_{16}-5_{23}$ transition at 22.23508 GHz)  of NGC 1068 reported here were made on 2000 February  23-24
 using the Very Long Baseline Array (VLBA), the phased Very Large Array (VLA) of the National Radio Astronomy Observatory (NRAO), 
 and the Effelsberg 100-m telescope at Bonn (PI: Greenhill, L. ). The $uv$ coverage is shown in figure 1. 
Eight IFs were recorded, each with a bandwidth of 8 MHz, divided into 512 channels (0.2 \kms velocity resolution). 
The range of the LSR velocities was between $690 - 1460$ \kms, covering systemic and high-velocity maser features (Claussen et al. 1984, Claussen \& Lo 1986, Nakai et al. 1995, Greenhill et al. 1996, Greenhill \& Gwinn 1997). 

The data were processed on the VLBA correlator at the NRAO. 
We used the observation coordinate of  ($\alpha$, $\delta$)$_{{\rm J}2000}$ = (2h 42m 40.70905s, $-$0$^\circ$ 0' 47.945")  as the phase tracking center. 
The correlated data were downloaded from the data archive system of the NRAO. 
The data reduction including both calibration and imaging were processed using the Astronomical Image Processing System (AIPS) package. 
Unfortunately, the data of the Mauna Kea station (MK) were removed from our analysis because of the low data quality due to bad weather conditions. 
The bandpass response was calibrated by observing 0133+476, and the residual delays and fringe rates were estimated using point-like sources  0237$-$027 and 0234+285.
The amplitude calibration was performed on the basis of the amplitude fluctuation of the point-sources 0133+476 and 0528+134 
whose flux densities were set to be 1.5 and 1.9 Jy, respectively.
The self calibration was applied using the strongest maser spot at $V_{\rm LSR}$ = 1174 \kms as a reference (IF 2). 
The obtained water maser spectrum of NGC 1068 is shown in figure 2. 
The total isotropic luminosity of the maser features was 50 $L_{\odot}$.

The imaging was performed with the CLEAN method. For all the detected maser spots, CLEAN  maps were made using an intermediate weighting between natural and uniform weighting. 
The maser spots were imaged by averaging 10 channels (2 \kms).  The synthesized beam  was 1.3 mas $\times$ 1.0 mas and PA=$-16$ deg.
The positions and intensity of the maser spots in each spectral channel were measured by using a  POPS pipeline script.
Brightness peaks higher than 4 $\sigma$ noise level (5 mJy/beam) through continuous two channels at least were identified as maser spots automatically and 
 their parameters as Gaussian brightness components were extracted. 

\section {Results and Discussion}

\subsection {The Observed Maser Features}

Figure 3(a) shows the distribution of the masers at the components S1 and C, together with the continuum map at 5 GHz obtained with MERLIN
 (Gallimore et al.  2004), at the coordinate system relative to the position of the maser spot near the systemic velocity of NGC 1068, { $V_{\rm LSR}=1137$ \kms.} 
The typical position errors were estimated from $\Delta \theta=\sqrt{\theta_{\rm fit}^2+\theta_{\rm rms}^2}$, where $\theta_{\rm fit}$ is the fitting error of
 2-dimensional Gaussian fit and  $\theta_{\rm rms} = 0.5~ \theta_{\rm beam}/{\rm SNR}$. 
No significant maser emission stronger than 4-$\sigma$ (5 mJy/beam) was detected out of the region of the S1 and C components in figure 3. 
The position and the velocity of the detected maser spots are shown in table 2 (The full table is available online).

The systemic, red- and blue-shifted maser spots were located at the  S1 component, which is thought to be the nucleus of the galaxy  (figure 3(b)). 
It is clear that the maser spots linearly extended from north-west to south-east, indicating the presence of the disk, with a position angle $=-54^\circ$,
 which differs from the galactic disk of NGC 1068 by $\sim 20^\circ$ (table 1). 
Other spots off the linear extensions distribute along the jet direction. 
The maser spots with  velocities around 900 \kms were also found at the  C component, which is thought to be the knot in the jet (figure 3(c)), 
showing a ring-like distribution pattern.

\subsection{The S1 Component}

\subsubsection{Rotating disk}
The linear distribution of the maser spots in the S1 component is well described by a thin and edge-on disk, already shown by Greenhill et al (1996) and Greenhill \& Gwinn (1997). 
Figure 4 shows a position-velocity diagram of these maser spots. 
The impact parameter is measured with respect to the strongest spot at $v_{\rm LSR}=1174 $ \kms.  
Assuming a circular rotating disk, we fitted the diagram to a linear rotation curve of $v\propto r$ for $r \leq r_{\rm in}$, and a Keplerian curve of 
$v\propto r^{-0.5}$ for $r_{\rm out}\geq r\geq r_{\rm in}$, where $v$ is the rotation velocity, $r$ is the distance from the center, $r_{\rm in}$ is the inner radius,
 and $r_{\rm out}$ is the outer radius of the maser disk. 
The fitting curve is also shown in figure 4, and we obtained the best fitting parameters of the central mass and disk based on the fitting results (table 3). 
The mass inside the radius of the rotating disk ($r_{\rm in}$=0.62 pc) is estimated to be $1.5 \times 10^7 M_{\odot}$, which is similar to the previous results 
(Gallimore et al. 1996a, Greenhill \& Gwinn 1997). 
Greenhill et al. (1996) indicated a sub-Keplerian curve ($v\propto r^{-0.31\pm0.02}$) of the masers in the outer portion of the disk (red-shifted component). 
In our data, at $r_{\rm in} < r < r_{\rm out}$, actually the best fit rotation curve is $v\propto r^{-0.24\pm0.10}$. 
The sub-Keplerian disk strongly indicate the effect of the self-gravity of the torus or circumnuclear stellar mass (Greenhill et al. 1996). 
Actually Hur{\'{e}} (2002) explained the sub-Keplerian rotation curve by the black hole mass of (1.2$\pm$0.1)$\times 10^7$ $M_{\odot}$ and 
the disk mass of  (9.4$\pm$1.6)$\times 10^6$ $M_{\odot}$  for the distance of 15 Mpc they adopted. 
Further analysis based on our high sensitive data will be made in our forthcoming paper. 

\subsubsection{Outflow}

In addition to the disk component, many weak maser spots can be found off the maser disk, within 20 mas (1.5 pc) from the central engine,  
lying close to the axis of the disk rotation (figures  3(b) and 5 (a)). 
These spots have not been detected in the previous observations. 
Greenhill et al. (2003) found the maser spots having similar properties at the Circinus Galaxy, and they inferred that they are associated with a wide-angle bipolar outflow. 
In figure 5(a),  most of maser spots at the north side are blue-shifted, indicating  the approaching outflow,
and those at the south are red-shifted,  indicating the counter outflow.

We show the schematic view in figure 5 (b). 
Most of these maser spots exhibiting blue-shifted are thought to be at the front of the outflow cones of continuum emission, and are amplified by the continuum radiation from the jets. 
The velocity difference between the bluest maser at the northern outflow and the reddest maser at the southern outflow is almost 600 \kms, 
indicating the apparent outflow speed of 300 \kms in the line of sight. This is slightly smaller than the molecular outflow speed of 450 \kms 
suggested from the blue wing in the HCN absorption feature (Impellizzeri et al. 2019).

  {Gallimore et al. (2016) found that velocity distribution of CO emission in the nucleus of NGC 1068 is consistent with the bipolar outflow with the speed of ~ 400 \kms in the direction nearly perpendicular to the maser disk. The position of the blue-shifted peak is displaced 1.2 pc northwest of the red-shifted peak. These properties found with CO observations are in good agreement with those in our VLBI observations. This fact indicated that the extended maser features in the S1 component are likely to exhibit the detailed structure of the wide-angle bipolar outflow on the sub-parsec scale.}

\subsection{The C Component}

Figure 3(c) shows the ring-like distribution of the maser spots at the C component.
Assuming that the maser spots around the C component distribute elliptically, we fitted them by a single ellipse.
We  summarized the fitted parameters in table 4. 
Some of the maser spots at the C component has been detected also with the VLA (Gallimore et al. 1996a), and are shown as the filled square in figure 6. 

Figure 6 (a)  shows an overlay of the maser emission on the continuum map at 5 GHz.
The center of the water maser distribution (cross) differs from the peak of the continuum emission by 12 mas  (0.92 pc).
The relative position between the distributions of the maser and continuum emission, however, may incorrect, because the adopted origins 
of the coordinate of the two emission at S1 are different (the maser spot at   { $V_{\rm LSR}=1133.2$ \kms} in the maser disk and the continuum peak
 in the ionized disk [Gallimore et al. 1997]). 
Thus we tried to shift the map of maser emission by 12 mas, so that  the maser center  coincides with the continuum peak. 
Figure 6 (b) shows the result, by which the maser spots look to surround the continuum emission. 

\subsubsection{Kinematics of the masers}

Figure 7 presents a position-velocity diagram cut along the major axis of the fitted ellipse shown in figure 6. 
We are hard to find some systematic tendency such as rotation and expansion from the very limited data in the position-velocity diagram. 
Then we introduced a position angle (PA) -velocity diagram along a fitted ellipse, adopting a simple model with a single ring which is rotating 
and expanding (or contracting) with constant speeds, $v_{\rm rot}$ and $v_{\rm exp}$, respectively (figure 8). 
We examined the radial velocity variation with the angle along the ellipse, $\omega$, which can be written
 as $\tan \omega = \cos \alpha \tan \theta$, where $\theta$ is the PA along a putative ring in the face-on view, and $\alpha$ is the viewing angle of the ring.
We used  $\alpha=44.6^\circ$ estimated from the ellipticity of the fitted ellipse. 
The observed radial velocity $v$ as a function of $\omega$ can be expressed as,

\begin{eqnarray}
v(\omega) = v_{\rm sys} + \sin \alpha \left( v_{\rm rot} \cos \theta(\omega) + v_{\rm exp} \sin  \theta(\omega)\right),  
\end{eqnarray}
where $v_{\rm sys}$ is the systemic velocity of the ring. 
Figure 9 shows the best fitting result by using equation (1). 
The obtained parameters are  $v_{\rm sys}=907\pm14$ \kms, $v_{\rm rot}=45\pm32$ \kms, and $v_{\rm exp}=117\pm28$ \kms, as shown by the solid line in figure 9. 
We also carried out fitting with other cases, considering only expansion or rotation, i.e., $v_{\rm rot}=0$ or  $v_{\rm exp}=0$, 
respectively, and also show the results in figure 9 by the dotted and dashed lines, respectively. 
While the rotation-only model is clearly not suitable for explaining the observational result,  
the expansion-only model is almost equivalent with the best fitted result by equation (1). 
These results are not enough significant, i.e., the obtained reduced $\chi^2$ lies between 5 $-$ 8 for the all cases, by adopting the error of the velocity of 2 \kms.   
However, it possibly imply that the expansion motion seems to more dominate the kinematics of the masers at the C component than the rotation motion. 

\subsubsection{The possible interpretation, jet collision or disk?}

Gallimore et al. (1996b, 2001) suggested that the jet masers at the C component indicate the presence of dense and warm molecular gas
 which lies on the radio jet emanating from the central engine at the S1 component. Gallimore et al. (2004) also discussed the jet-shocked model 
 based on the free-free opacity estimated from the spectrum energy distribution of the C component. 

Recent ALMA observations of NGC 1068 showed that high-density molecular gas tracers HCN and HCO$^+$ gases almost coincide to the C component (Imanishi et al. 2016, 2018). 
Figure 10 shows the positional relationship between the molecular gas  and the S1 and C components. 
The radial velocity of HCN and HCO$^+$ at the C component is 1060 and  1080 \kms in the radio definition, respectively. 
The averaged velocity of the water maser at the C component is  $\sim900$ \kms, showing that velocities of HCN and HCO$^+$ are faster than  
the maser velocity by $\sim170$ \kms. 

Considering the presence of the molecular cloud on the jet shown by the ALMA observations, we can interpret that the radio jet collides with 
the molecular cloud, then the shock wave due to the jet collision induced an ionized region in the molecular cloud. At the circumstance of
 the expanding ionized sphere, the water masers could be excited by the shock collisional  pumping. 
Such a jet collision picture has also been proposed in the jet of the Seyfert NGC 3079, exhibited by the result of the evolution of the knot in the jet (Middelberg et al. 2007). 

On the other hand, the ring-like distribution of the maser spots in figure 3 (c) admits of another interpretation, a face-on disk. 
In this picture, the C component  itself would be an AGN, i.e., a SMBH, because a megamaser exists only in an AGN, implying that NGC 1068 has a binary of SMBH. 
This is not curious because NGC 1068 shows several lines of evidence for a past minor merger event several billion years ago
 (see Tanaka et al. 2017 and references therein). If the merging satellite galaxy has a SMBH at its center, one may observe a binary of SMBHs
  in the final phase of the merger (e.g., Begelman et al. 1986).  

Such a face-on maser disk is needed to be geometrically thick, because the path length should be long enough to amplify the maser emission effectively. 
Although generally the disk maser emission is likely to be amplified along the limb of the disk, e.g., NGC 4258 (Miyoshi et al. 1995), 
a relatively thick disk around the SMBH has been seen in the maser disks such as the Seyfert NGC 3079 (e.g., Yamauchi et al. 2004, Kondratko et al. 2005) 
and IC 1481 (Mamyoda et al. 2009). 

If the C component really has a secondary SMBH of NGC 1068, HCN and HCO$^+$ emission found by ALMA might be the secondary dusty molecular torus. 
It is interesting that a pair of compact jet-like feature is emanated from the component C toward NW and SE directions in the VLBA 5 GHz image
 (Gallimore et al. 2004, see also figure 6); 
each of the compact jet-like feature has a projected length of $\sim$ 30 mas, corresponding to $\sim$ 2.3 pc.
The origin of the molecular cloud at the C component is a very important problem to be solved by further multi frequency studies. 

If the SMBH hypothesis is correct, the enclosed mass in the ring can be estimated to be $9\times10^5 M_{\odot}$ by adopting the rotation speed of 
45 \kms from the ring radius of 2 pc. 
In order to confirm whether or not the component C really has a SMBH, we need to detect more maser spots at the C component and clarify their motion, 
and estimate the mass containing in the molecular cloud by using the gas kinematics.

\section{Conclusion}

We imaged the water maser emission at 22 GHz at the nuclear radio component S1 and the jet component C in the Seyfert galaxy NGC 1068 
with the high sensitive VLBI observations.
The main results  are summarized as follows.

1. At the S1 component, the water maser spots align with a maser disk. The dynamical mass enclosed within a radius of 0.62 pc is estimated t
o be $1.5\times 10^7$ $M_{\odot}$.  This is consistent with those obtained with the previous VLBA observations (Gallimore et al. 1996a, Greenhill et al. 1996). 
The best fit rotation curve showed sub-Keplerian, $v\propto r^{-0.24\pm0.10}$, indicating  the self-gravity of the molecular disk needed to be considered. 

2. In addition to the maser disk, the distribution of the blue- and red-shifted components of the water maser spots perpendicular to the disk was detected, 
suggesting the presence of a wide-angle outflow of the water maser extending to 1.5 pc and having the speed of $\sim300$ \kms.

3. At the C component, the ring-like distribution of the maser spots was detected for the first time. The diameter of the ring is  about 22 mas (1.7 pc).  
These masers might expand or contract with the speed of $\sim 120$ \kms.  

4. Since the C component well coincides with the HCN and HCO$^+$ molecular cloud  found with ALMA, the ring-like structure can be considered to be 
caused by the jet collision to the molecular cloud which has been proposed by Gallimore et al. (1996b, 2001). 

5. Another possible explanation is the face-on disk at the C component. This might suggest the C component has a SMBH, implying the presence of 
a binary SMBH in NGC 1068. This idea seems reasonable because NGC 1068 shows several lines of evidence for a past minor merger event (Tanaka et al. 2017).

\begin{ack}

  {We would like to thank the anonymous referees, who
gave us many useful comments that improved the paper very much.}
We would like to thank Drs. Hiroshi Imai, Satoko Sawada-Sato, Yoshiaki Hagiwara, and Akihiro Doi for their kind supports to data reduction, imaging, and interpretation of VLBI data. We also grateful to Dr. Masatoshi Imanishi for providing the ALMA image of NGC 1068. We gratefully acknowledge the staff at the NRAO VLBA for their invaluable help both in providing online archive and in data reduction.

\end{ack}

%%%%%%%%%%%%%%%%%%%%figure 1
\begin{figure}
 \begin{center}
  \includegraphics[width=15cm]{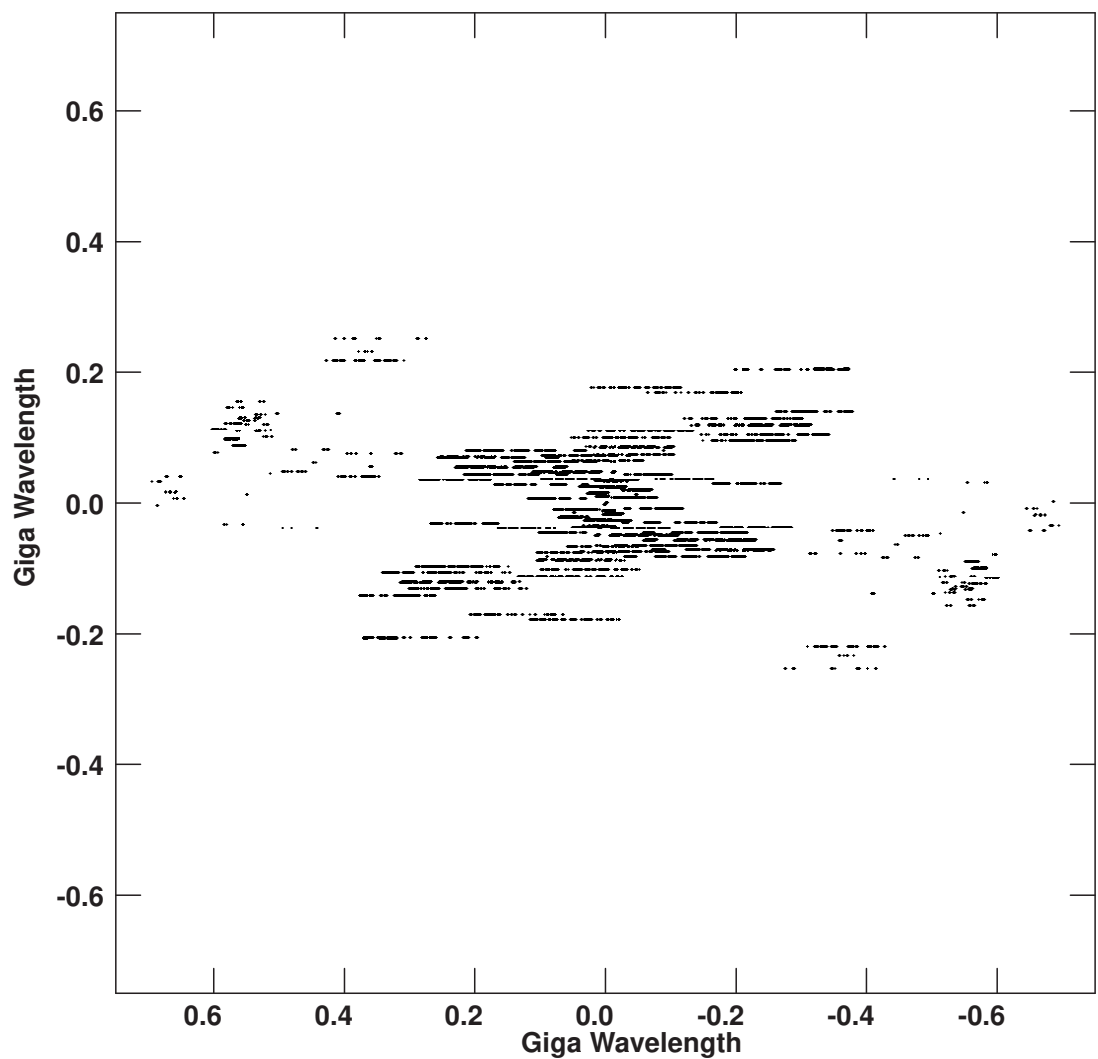}
 \end{center} \vspace{4cm}
 \caption{The u-v couverage of our data. }\label{fig1}
\end{figure}
\clearpage

%%%%%%%%%%%%%%%%%%%%figure 2
\begin{figure}
 \begin{center}
  \includegraphics[width=12 cm]{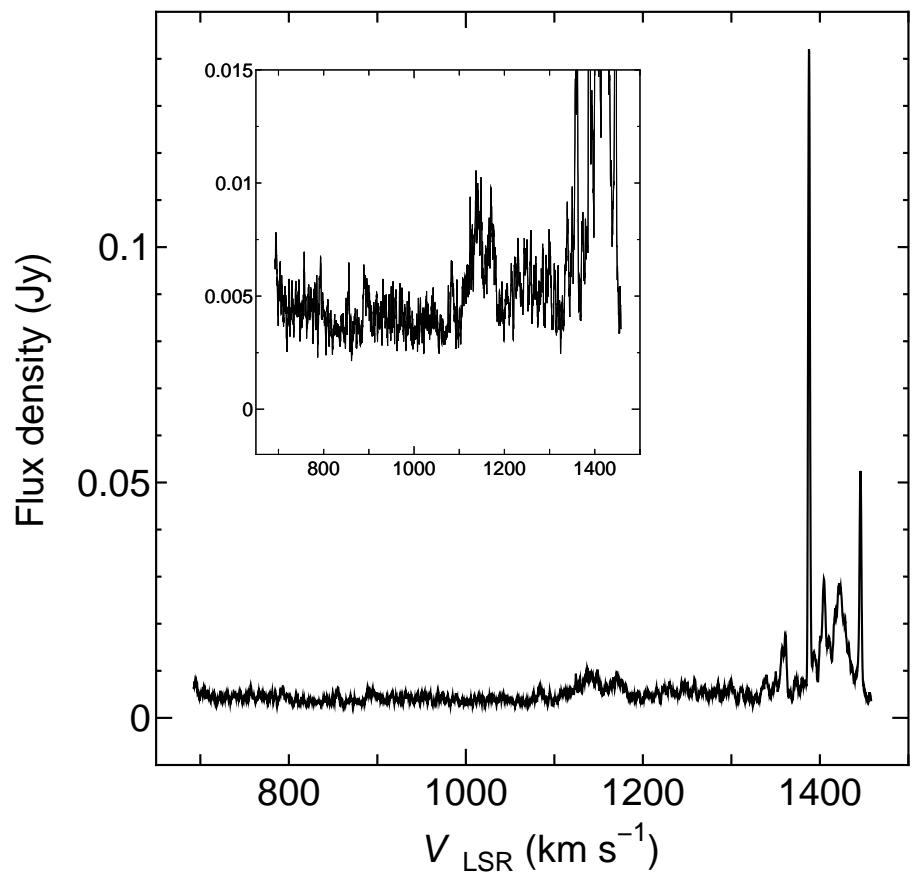}
 \end{center} \vspace{4cm}
 \caption{The water maser spectrum. }\label{fig2}
\end{figure}
\clearpage
%%%%%%%%%%%%%%%%%%%%figure 3
\begin{figure}
 \begin{center}
  \includegraphics[width=12 cm]{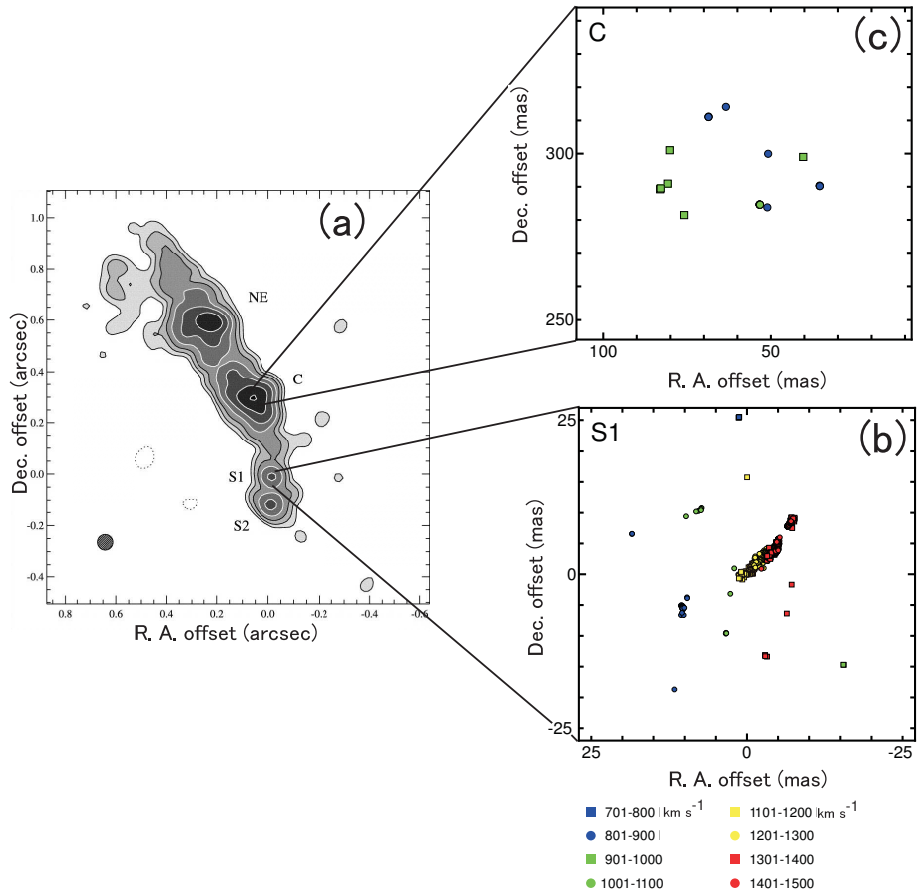}
 \end{center} \vspace{2cm}
 \caption{(a) MERLIN 5 GHz continuum image in contours (Gallimore et al. 2004). (b) The distribution of  maser spots in the S1 component. (c) The distribution of maser spots at the C component.
The color scale  gives the LSR velocity. The coordinate system of the map of the maser spots is relative to the position of the maser spot   { near the systemic velocity of NGC 1068, $V_{\rm LSR}=1137$ \kms,} in the S1 component. }\label{fig3}
\end{figure}
\clearpage

%%%%%%%%%%%%%%%%%%%%figure 4
\begin{figure}
 \begin{center}
  \includegraphics[width=9cm]{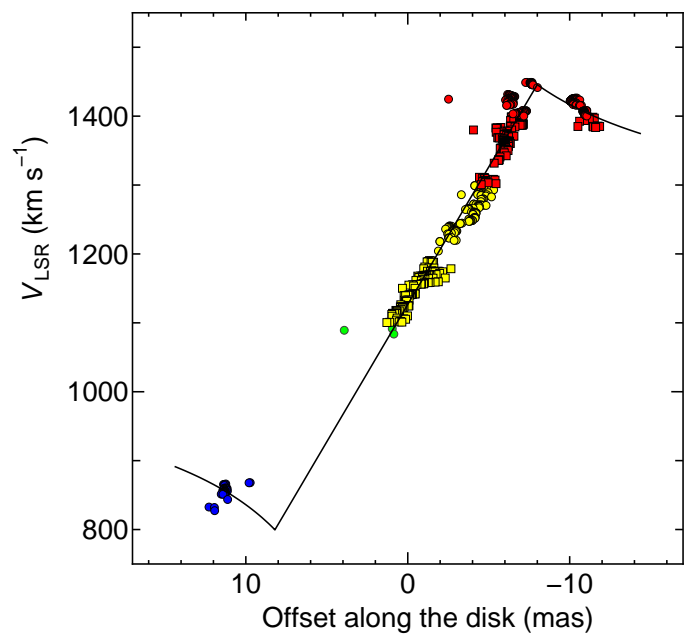}
 \end{center} \vspace{4cm}
 \caption{The position-velocity diagram of the water maser spots at the Component S1. The abscissa is the offset from the maser spot of $V_{\rm LSR}=1174$ \kms along the maser disk with PA$=286^\circ$. The color scale give the LSR velocity (see figure 3). The solid line indicates the Keplerian fit.} %%%and the dotted line indicates the best fit result of $v\propto r^{-0.22}$.}
\label{fig4}
\end{figure}
\clearpage

%%%%%%%%%%%%%%%%%%%%figure 5
\begin{figure}
 \begin{center}
 \includegraphics[width=12 cm]{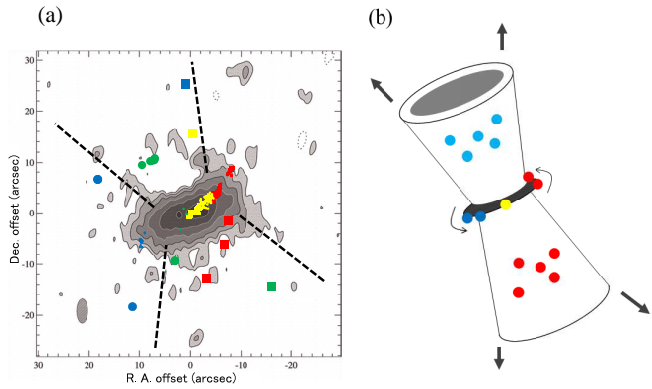}
 \end{center} \vspace{4cm}
 \caption{(a)  The water masers in the S1 component relative to the VLBA 5 GHz continuum images in contours (Gallimore et al. 2004). The dotted line indicates the limb of the expected  wide-angle outflow. (b) Schematic view for the disk and outflow model of the component S1. The blue, red, and green circles indicate the blue-shifted, red-shifted, and systemic velocity components. The coordinate system of the map of the maser spots is relative to the position of the maser spot near the systemic velocity of NGC 1068,   { $V_{\rm LSR}=1137$ \kms.}  }\label{fig5}
\end{figure}
\clearpage

%%%%%%%%%%%%%%%%%%%%figure 6
\begin{figure}
 \begin{center}
  \includegraphics[width=12cm]{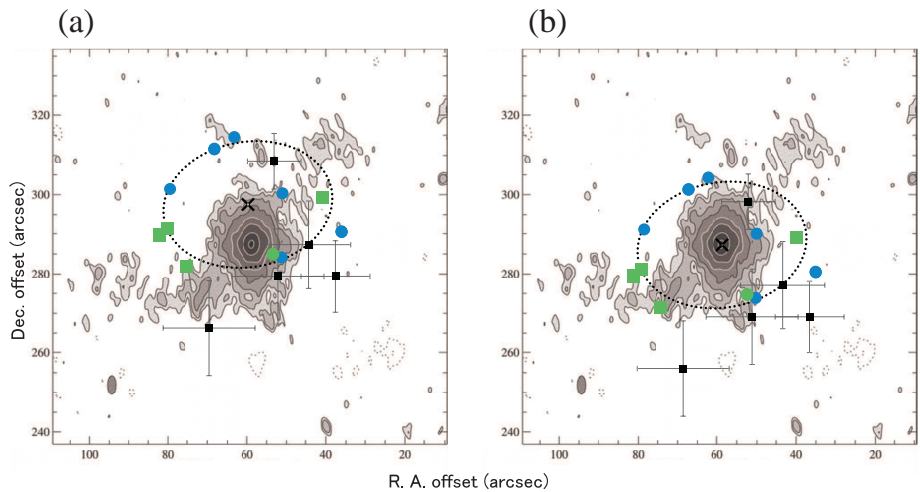}
 \end{center} \vspace{4cm}
 \caption{ (a) Overlay of the water masers at the C component on the VLBA 5 GHz continuum images in contours (Gallimore et al. 2004). The coordinate system of the map of the maser spots is relative to the position of the maser spot near the systemic velocity of NGC 1068,   { $V_{\rm LSR}=1137$ \kms.} The relative position between the maser and 5-GHz continuum is determined, assuming that the rotation center of the maser disk at the S1 (i.e., galactic center) in figure 4 coincides with the peak position of the 5-GHz continuum peak. (b) Same as (a), but the map of the water maser is shifted so that the center of the maser distribution (cross) and the continuum peak coincide. Filled circles indicate the maser spots in our results, and filled squares indicate  those in Gallimore et al. (1996a). The color scale give the LSR radial velocity (see figure 3). The dotted line indicates the best fit ellipse (see also table 3). }\label{fig6}
\end{figure}
\clearpage

%%%%%%%%%%%%%%%%%%%%figure 7
\begin{figure}
 \begin{center}
  \includegraphics[width=8cm]{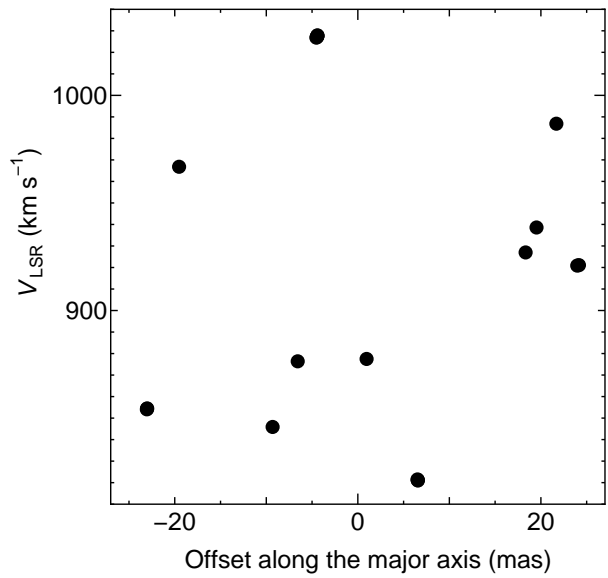}
 \end{center} \vspace{4cm}
 \caption{Position-velocity diagram of the maser spots at the C component. The positions were measured along the major axis of the fitted ellipse (PA=286$^\circ$) shown in figure 6. }\label{fig8}
\end{figure}
\clearpage

%%%%%%%%%%%%%%%%%%%%figure 8
\begin{figure}
 \begin{center}
  \includegraphics[width=12 cm]{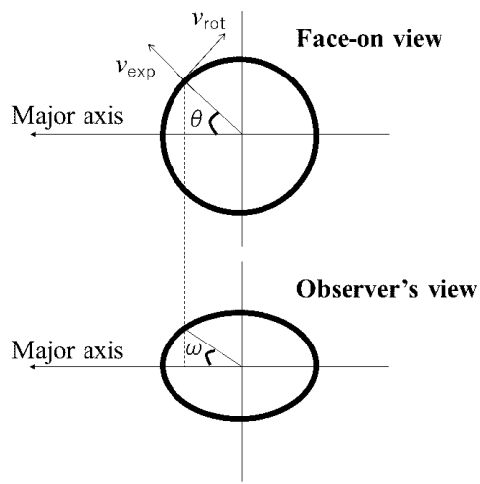}
 \end{center} \vspace{4cm}
 \caption{A ring model of the kinematics of the maser spots at the C component.  
The top panel shows the face-on view. The maser spots distribute on the ring rotating and expanding with the constant speeds. $\theta$ indicates the position angle. The bottom panel shows the observer's view. The apparent position angle, $\omega$, is likely to be smaller than $\theta$, due to the geometrical effect. }\label{fig8}
\end{figure}
\clearpage

%%%%%%%%%%%%%%%%%%%%figure 9
\begin{figure}
 \begin{center}
  \includegraphics[width=8cm]{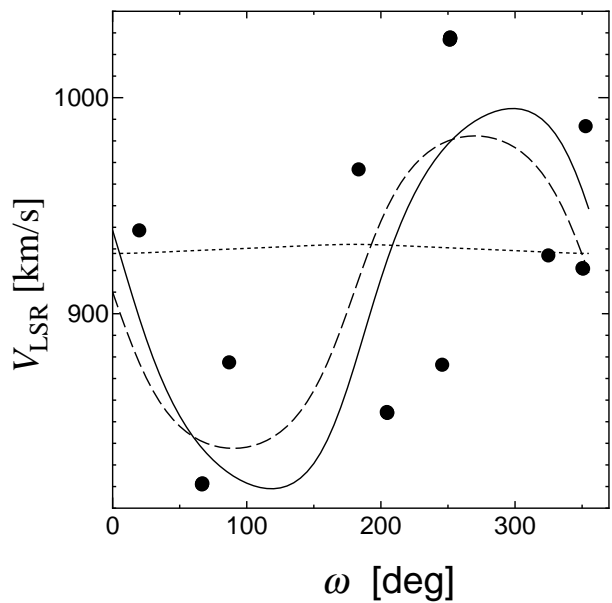}
 \end{center} \vspace{4cm}
 \caption{The fitting results of the position angle-velocity diagram of the maser spots at the C component, based on the model shown in figure 8. The solid line indicates the best fit result. 
%%%% and the fitted parameters are shown in table 5. 
The dotted line and dashed line indicate the results at the Case of $v_{\rm exp}=0$ and $v_{\rm rot}=0$, respectively. 
%%%Note that these results are not significant, exhibiting the reduced $\chi^2$ ranges between 5 -- 8. 
}\label{fig8}
\end{figure}
\clearpage

%%%%%%%%%%%%%%%%%%%%figure 10
\begin{figure}
 \begin{center}
  \includegraphics[width=12cm]{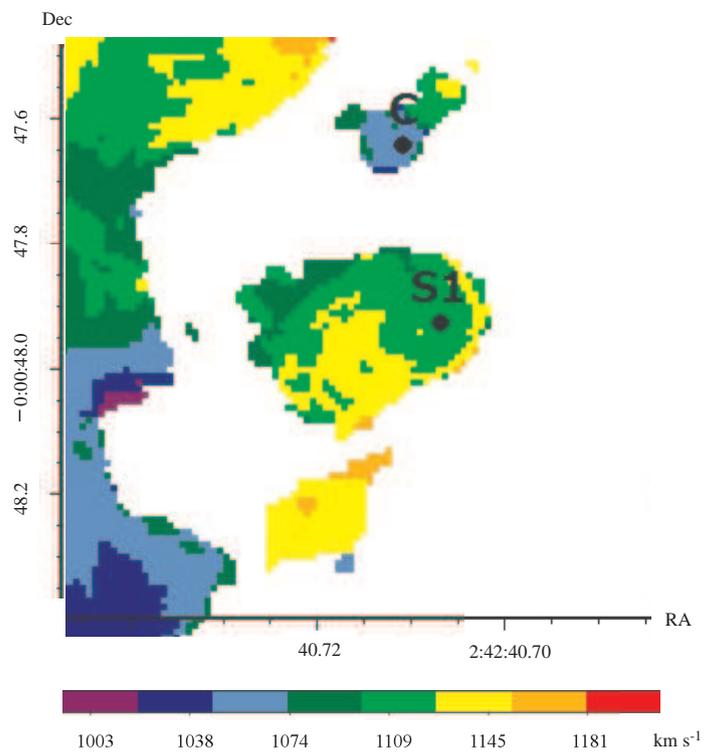}
 \end{center} \vspace{2cm}
 \caption{The HCN map obtained with ALMA (Imanishi et al. 2018). The color scale gives the LSR radial velocity in the optical definition. The relationship of the velocities between radio definition and optical definition  is $v_{\rm radio}=v_{\rm optical}/(1+z)$. For NGC 1068, since  $z=0.00379$, $v_{\rm radio}=0.996 v_{\rm optical}$}.\label{fig7}
\end{figure}
\clearpage

%%%%%%%%%%%%%%%%%%%%%%%%%%%%%%%%%%%%%%%%%%%%%%%%%%%%%%%%%%%%%%%%%%%%%%%
%% Table 1 : adopted parameters %%%%%%%%%%%%%%%%%%%%%%%%%%%%%%%%%%%%%%%
%%%%%%%%%%%%%%%%%%%%%%%%%%%%%%%%%%%%%%%%%%%%%%%%%%%%%%%%%%%%%%%%%%%%%%%
\begin{table}[htbp]
  \tbl{Adopted parameters of NGC 1068.}{%
  \begin{tabular}{ll}
     \hline
Phase tracking center	   & $\alpha_{2000} = \timeform{2h42m40s.70905}$ \\
					   & $\delta_{2000} = \timeform{-0D0'47''.945}$ \\
Morphological type\footnotemark[$\dagger$] & (R)SA(rs)b   \\
Distance\footnotemark[$\ddagger$]	   & 14 Mpc \\
Systemic velocity\footnotemark[$\S$]	   & 1137 km s$^{-1}$ \\
Inclination angle\footnotemark[$\|$]  & $40^{\circ} \pm 3^{\circ}$  \\
Position angle\footnotemark[$\|$]	   & $286^{\circ} \pm 5^{\circ}$ \\
\hline
    \end{tabular}}\label{adopt}
\begin{tabnote}
\footnotemark[$\dagger$] NED. \\
\footnotemark[$\ddagger$] Tully \& Fisher 1988. \\
\footnotemark[$\S$] LSR velocity converted from heliocentric velocity by Brinks \etal\ 1997. \\
\footnotemark[$\|$] Brinks \etal\ 1997. \\
\end{tabnote}
\end{table}

%%%%%%%%%%%%%table 2

\begin{deluxetable}{rrrrrrrr} 
\tablecaption{The positions and velocities of the maser spots in NGC 1068. The full table is available online. }

\tabletypesize{\footnotesize}
\tablewidth{0pt}

\tablehead{
\colhead{Velocity}	&	\colhead{R.A. Offset}			&	\colhead{Dec. Offset}			&	\colhead{Peak int.}			&	\colhead{Flux}		\\

\colhead{[kms$^{-1}$]}  &   \colhead{[mas]}  &  \colhead{ [mas]}    &   \colhead{ [mJy/beam]}    &   \colhead{[mJy]}   

}

\startdata

{\bf S1: Disk}\\
\hline
718.4 	&	-25.4 	$\pm$	0.3 	&	8.6 	$\pm$	0.3 	&	3.8 	$\pm$	0.8 	&	19.7 	$\pm$	4.9 	&	\\
719.1 	&	-24.7 	$\pm$	0.1 	&	-8.3 	$\pm$	0.1 	&	4.9 	$\pm$	0.9 	&	4.4 	$\pm$	1.4 	&	\\
723.7 	&	-3.8 	$\pm$	0.3 	&	14.3 	$\pm$	0.3 	&	3.5 	$\pm$	0.8 	&	18.8 	$\pm$	4.8 	\\
731.1 	&	3.0 	$\pm$	0.1 	&	23.2 	$\pm$	0.1 	&	6.7 	$\pm$	0.8 	&	6.6 	$\pm$	1.5 	\\
735.9 	&	-3.6 	$\pm$	0.1 	&	-10.9 	$\pm$	0.1 	&	11.4 	$\pm$	0.9 	&	22.1 	$\pm$	2.5 	\\

\hline

\enddata
%%%\end{tabular} 
%%%\tablecomments{
%%%}
\end{deluxetable}

%%%%%%%%%%%%%table 3

\begin{table}[htbp]
\caption{Parameters of the maser disk at the Component S1.} 
\label{tab:p} 
\begin{center}
\begin{tabular}{rc} 
\hline
\hline
$r_{\rm in}$  [pc]& 0.62 (8.1 mas)  \\
$r_{\rm out}$ [pc]&  0.92 (11.9 mas)  \\
$v_{\rm in}$  [\kms]&  320 \\
$v_{\rm out}$ [\kms]& 280   \\
$M$ & $1.5 \times 10^7$ $M_{\odot}$ \\
\hline
\end{tabular} 
\begin{tabnote}
$r_{\rm in}$ is the inner radius, $r_{\rm out}$ is the outer radius, $v_{\rm in}$ is the inner velocity, $v_{\rm out}$ is the outer velocity, and $M$ is the mass of the SMBH. 
\end{tabnote}
\end{center} 
\end{table} 

%%%%%%%%%%%%%table 4

\begin{table}[htbp]
\caption{Elliptical fitting result at the C component.} 
\label{tab:e} 
\begin{center}
\begin{tabular}{lc} 
\hline
\hline
Center Position & \\
~ ~ ~ R.A. offset [mas] &$59.9\pm0.2$ ($4.60\pm0.02$ pc)\\
~ ~ ~ Dec. offset [mas] & $297.1\pm0.2$ ($22.84\pm0.02$ pc)\\
Major axis [mas] & $22.18\pm0.06$ ($1.706\pm0.004$ pc)\\
Minor axis [mas]&$15.80\pm0.04$ ($1.215\pm0.003$ pc)\\
Position angle [deg] & $-9\pm1$\\
\hline
\end{tabular} 
\begin{tabnote}
\end{tabnote}
\end{center} 
\end{table} 

\end{document}